\documentclass[aps,prl,twocolumn,showpacs]{revtex4-1}
\usepackage{graphicx,bm,braket}
\usepackage{natbib,ulem}
\begin{document}

\title{Exceptional Suppression of Flux-Flow Resistivity in FeSe$_{0.4}$Te$_{0.6}$ by Back-Flow from Excess Fe Atoms and Se/Te Substitutions}

\author{
Tatsunori Okada, Fuyuki Nabeshima, Hideyuki Takahashi, Yoshinori Imai, and Atsutaka Maeda
}
\affiliation{
Department of Basic Science, The University of Tokyo, Meguro-ku, Tokyo 153-8902, Japan
}

\date{\today}

\begin{abstract}
We measured the microwave surface impedance of FeSe$_{0.4}$Te$_{0.6}$ single crystals with- and without external magnetic fields.
The superfluid density exhibited a quadratic temperature dependence, indicating a strong pair-breaking effect.
The flux-flow resistivity behaved as $\rho_f(B\ll B_{\rm c2})/\rho_n=\alpha B/B_{\rm c2}$.
The observed $\alpha$ value of $\approx 0.66$ was considerably smaller than that of other Fe-based materials ($\alpha\geq1$) and was attributed to a back-flow of superfluids remarkable in disordered superconductors.
This is the first-time observation of the back-flow phenomenon caused by an origin other than the vortex pinning in multiple-band systems.
\end{abstract}
\maketitle
\section{I. Introduction}
Following the discovery of superconductivity in LaFeAsO$_{1-x}$F$_{x}$ \cite{Kamihara2008}, Fe-based superconductors (Fe-SCs) have been extensively investigated worldwide.
Fe-SCs exhibit multiple bands/gaps: thus, it has been predicted that the superconducting order parameter could change its sign among different sheets of the Fermi surface \cite{Mazin2008,Kuroki2008}, and various gap structures have been observed \cite{Hirschfeld2011}.
To elucidate the mechanism of such novel SCs, the gap structure of each material should be systematically investigated, and essential characteristics of the Fe-SCs should be extracted from the accumulated data.

In addition to conventional probes that are sensitive to low-energy excitations, such as the temperature-dependent magnetic penetration depth $\lambda(T)$, the magnetic field dependence of the flux-flow resistivity, $\rho_{f}(B)$, is known to be sensitive to the superconducting gap structure since $\rho_f$ is induced by quasiparticles excited inside the vortex core reflecting the gap function.
For most SCs, $\rho_f(B)$ at low fields behaves as $\rho_f(B)/\rho_n\approx\alpha B/B_{\rm c2}$, where $\rho_{n}$ and $B_{\rm c2}$ are the normal-state resistivity and the upper critical field, respectively.
The structure of the superconducting gap is reflected in the gradient $\alpha$.
Specifically, $\alpha$ values of conventional SCs with an isotropic gap are almost unity \cite{Strnad1964}, which are explained by the Bardeen-Stephen (B-S) theory \cite{Bardeen1965}.
In contrast, unconventional SCs with $p$-wave \cite{Kambe1999}, $d$-wave \cite{Tsuchiya2001, Matsuda2002}, and anisotropic $s$-wave \cite{Takaki2002} symmetry exhibit $\alpha$s above unity.
Kopnin and Volovik (K-V) \cite{Kopnin1997} justified the empirical relationship in which $\alpha$ increases with the anisotropy of the gap function by accounting for bound states inside the vortex core.
Large $\alpha$s have also been found in two-band SCs \cite{Shibata2003,Akutagawa2008,Goryo2005}.

Novel phenomena have been predicted for multiple-band SCs such as the dissociation of a flux line into a couple of fractional flux quantum \cite{Lin2013} and the time-reversal-symmetry-breaking state \cite{Hu2012}.
Thus, it is both interesting and significant to experimentally investigate characteristics of vortices in multiple-band SCs.
To determine how novel features of Fe-SCs appear in the flux-flow state, thus far, we have investigated the $\rho_f(B)$ of several Fe-based materials, such as LiFeAs (Li111) \cite{Okada2012_Li111}, LiFeAs$_{0.97}$P$_{0.03}$ (P-Li111) \cite{Okada2013_P-Li111}, NaFe$_{0.97}$Co$_{0.03}$As (Co-Na111) \cite{Okada2013_Co-Na111}, SrFe$_{2}$(As$_{0.7}$P$_{0.3}$)$_{2}$ (P-Sr122) \cite{Takahashi2012_P-Sr122}, and BaFe$_2$(As$_{0.55}$P$_{0.45}$)$_2$ (P-Ba122) \cite{Okada2014_P-Ba122}.
The primary contributions of these studies were that (i) observed $\alpha$ values are significantly different from each other and (i\hspace{-0.1em}i) $\alpha$ tends to increase when at least one highly anisotropic gap is present, which is somewhat similar to the behavior in single-band SCs.
We recently confirmed this tendency in Li111 and P-Ba122 by quantitatively evaluating a relation between $\alpha$ and the gap anisotropy by extending the K-V model to two-band systems \cite{Okada_future}.
Based on those systematic studies for $\rho_f(B)$ of Fe-SCs, the gap-anisotropy scenario is probably common to all of the Fe-SCs.
However, $\rho_f(B)$ of Fe-SCs with strong impurity scattering remains unclear because existing flux-flow data for Fe-SCs have mostly been obtained for fairly clean materials, and there is no theoretical research as for the effect of strong disorder on vortices of multiple-band SCs.
Although we have already clarified that Co-Na111 exhibits gapless superconductivity, we have not elucidated a relation between $\alpha$ and the amount/strength of impurities.
To elucidate the role of impurity scattering for $\rho_f(B)$, we focused on the FeSe$_{1-x}$Te$_{x}$ system.
It is well known that excess-Fe atoms enter Fe-(I\hspace{-0.1em}I) sites easily and act as magnetic impurities \cite{Komiya2013,Zhang2009}.
Therefore, FeSe$_{1-x}$Te$_{x}$ is an appropriate materials for investigating $\rho_f(B)$ of Fe-SCs with strong impurity scattering.

In this paper, we report on microwave surface impedance measurements of FeSe$_{0.4}$Te$_{0.6}$ single crystals both in the zero-field limit and under finite magnetic fields.
Observed results for $\lambda(T)$ and a parameter regarding a vortex pinning indicated that FeSe$_{0.4}$Te$_{0.6}$ was a SC in the dirty limit.
We also observed that $\alpha$ of this material was exceptionally small because of considerable back-flow current that was generated in SCs with disorder.
\section{I\hspace{-0.1em}I. Experiment}
Single crystals of FeSe$_{1-x}$Te$_{x}$ were grown using a method described elsewhere \cite{Taen2009, Noji2010}.
A composition analysis using energy dispersive X-ray spectroscopy (EDX) was performed on samples with a nominal composition of $\rm{Fe}:\rm{Se}:\rm{Te}=1:0.4:0.6$.
The corresponding actual ratios were found to be $1.00\pm0.04:0.37\pm0.05:0.63\pm0.02$.
Henceforth, we denote this composition by FeSe$_{0.4}$Te$_{0.6}$.
We confirmed the reproducibility of the results described in this paper by measuring four specimens cut from different batches of single crystals.

\begin{figure}[h]
	\includegraphics[width=0.9\hsize]{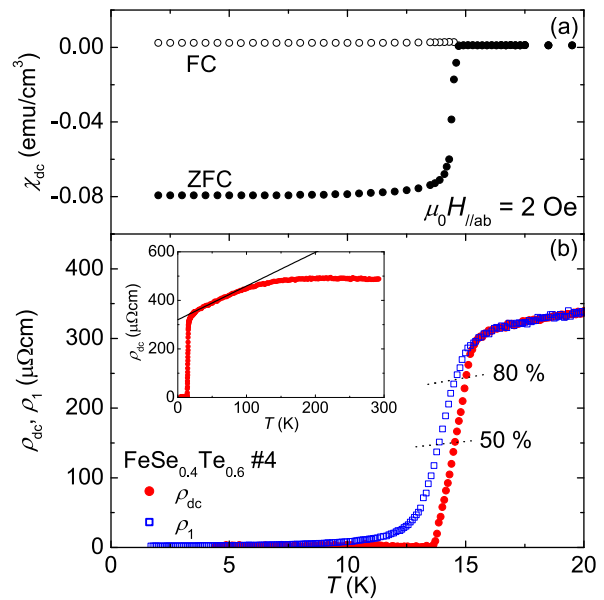}
	\caption{\label{fig:Tc}(Color online) The temperature dependence of electric- and magnetic properties of FeSe$_{0.4}$Te$_{0.6}$ batch \#4.
	(a) The dc magnetic susceptibility with both zero-field-cooled (ZFC) and field-cooled (FC) conditions under 2 Oe applied parallel to the $ab$-plane.
	(b) The dc resistivity (red circle) and the real part of the complex resistivity (blue square).
	Dotted lines are 50\% and 80\% of $\rho_n$.
	The inset shows $\rho_{\rm dc}$ up to room temperature and the extrapolation line of the linear part of $\rho_n$.
	}
\end{figure}

Figure \ref{fig:Tc}(a) shows the dc magnetic susceptibility as a function of temperature, $\chi_{\rm dc}(T)$, measured by a superconducting quantum interference device (SQUID) magnetometer.
$\chi_{\rm dc}(T)$ indicated a bulk superconductivity of $T_{\rm c}=14.6\ {\rm K}$.
Figure \ref{fig:Tc}(b) shows the temperature-dependent dc resistivity, $\rho_{\rm{dc}}(T)$, which was measured using a four-probe method.
The temperature where $\rho_{\rm dc}(T)$ drops to 50\% of the normal-state resistivity, $\rho_n(T)$, obtained by extrapolating $\rho_{\rm{dc}}(T)$ linearly to the superconducting region (shown as the solid line in the inset of Fig. \ref{fig:Tc}(b)) was $14.6\ \rm{K}$.
The residual resistivity of $\rho_n(0)=300\pm25\ \rm{\mu\Omega cm}$ is consistent with our previous report \cite{Nabeshima2012} and much larger than that of clean Fe-SCs such as Li111 ($\approx30\ \rm{\mu\Omega cm}$) and P-Sr122 ($\approx50\ \rm{\mu\Omega cm}$), indicating a strong impurity scattering in this material.
To measure the surface impedance, single crystals were cut into a small piece with typical dimensions of $a\times b\times c=0.5\times0.5\times0.2\ \rm{mm}^{3}$.

The microwave surface impedance $Z_{\rm s}=R_{\rm s}-{\rm i}X_{\rm s}$, where $R_{\rm{s}}$ and $X_{\rm{s}}$ denote the surface resistance and the surface reactance, was measured using cavity perturbation technique \cite{Maeda2005} with a cylindrical oxygen-free-Cu cavity resonator operated in the TE$_{011}$ mode.
The resonant frequency and the quality factor of the resonator, and the filling factor of the sample were $\omega/2\pi\approx19\ \rm{GHz}$, $Q\gtrsim6\times10^{4}$, and $F\approx6\times10^{-6}$, respectively.
Both an external field, $B=0-8\ {\rm T}$, and a microwave field, $B_{\omega}$, were applied parallel to the $c$-axis of the sample (a schematic is shown in the inset of Fig. \ref{fig:lambdavsT}).
The magnitude of $Z_{\rm s}$ was determined by assuming the Hagen-Rubens limit in the normal state; $R_{\rm s}=X_{\rm s}=\sqrt{\mu_0\omega\rho_{\rm dc}/2}$ ($\mu_0$: the vacuum permeability).
The details of this procedure are described elsewhere \cite{Matsuda2002,Maeda2005,Okada2012_Li111,Takahashi2012_P-Sr122}.
The real part of the complex resistivity, $\rho_1-{\rm i}\rho_2={\rm i}Z_{\rm s}^2/\mu_0\omega$, calculated from the measured $Z_{\rm s}$ is shown in Fig. \ref{fig:Tc}(b): the temperature at which $\rho_1$ becomes 80\% of $\rho_n$ corresponded to transition temperatures appeared in the data of $\chi_{\rm dc}(T)$ and $\rho_{\rm dc}(T)$.
Thus, we used the criteria of $\rho_1=0.8\rho_n$ to determine $T_{\rm c}$ from the measured $Z_{\rm s}(T,B)$.

We analyzed the flux-flow resistivity using the Coffey-Clem model, where $Z_{\rm s}$ induced by the vortex motion is calculated \cite{Coffey1991}.
The flux creep and the thermal fluctuations are negligibly small at sufficiently low temperatures; the Coffey-Clem model leads to a relation
\begin{equation}
	Z_{\rm{s}}=-{\rm i}\mu_{0}\omega\lambda\sqrt{1+{\rm i}\frac{\rho_{f}}{\mu_{0}\omega\lambda^{2}}\left(1-{\rm i}\frac{\omega_{\rm cr}}{\omega}\right)^{-1}},
\end{equation}
where $\omega_{\rm{cr}}/2\pi$ is the crossover frequency that characterized the crossover between the resistive response ($\omega>\omega_{\rm{cr}}$) and the reactive response ($\omega<\omega_{\rm{cr}}$).
Consequently, at $T\ll T_{\rm c}$, we could directly obtain $\rho_{f}(T,B)$, $\omega_{\rm{cr}}(T,B)$, and $\lambda(T,0)=X_{\rm{s}}(T,0)/\mu_{0}\omega$ from $R_{\rm{s}}(T,B)$ and $X_{\rm{s}}(T,B)$.
\section{I\hspace{-0.1em}I\hspace{-0.1em}I. Results and discussion}
\begin{figure}[h]
	\includegraphics[width=0.9\hsize]{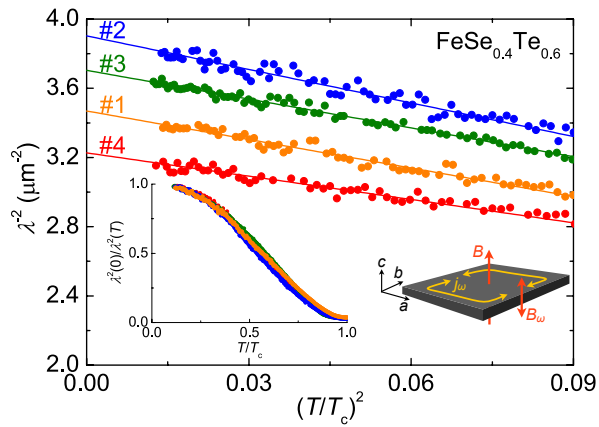}
	\caption{\label{fig:lambdavsT}(Color online) $\lambda^{-2}$ of FeSe$_{0.4}$Te$_{0.6}$ as a function of $(T/T_{\rm c})^2$ measured with $B=0\ \rm{T}$.
	Symbols are the data of batches \#1 (orange), \#2 (blue), \#3 (green), and \#4 (red), and solid lines are results fitted by a function $\lambda^{-2}(T)=\lambda^{-2}(0)[1-A(T/T_{\rm c})^{2}]$ below $0.3T_{\rm c}$.
	Insets show the $T$-dependent superfluid-density fraction $\lambda^2(0)/\lambda^2(T)$ (left) and the configuration of our experiment (right).
	}
\end{figure}

Figure \ref{fig:lambdavsT} shows the temperature dependence of $\lambda^{-2}$, which is proportional to the superfluid density, obtained from the data taken in the zero-field limit.
It can be clearly seen that $\lambda^{-2}(T)$ changed as $\lambda^{-2}(0)[1-A(T/T_{\rm c})^{n}]$ with an exponent of $n\approx2$, and both $\lambda(0)$ and $A$ determined by fitting the data with this function are listed in Table \ref{tab:sample properties}.
The two-dimensionality of the Fermi surface makes an existence of point nodes unlikely in FeSe$_{0.4}$Te$_{0.6}$.
Thus, the $T^2$-dependence shows that gapless superconductivity was induced by the pair-breaking effect in this material.
The results of the $T^{2}$-dependence and $\lambda(0)=530\pm27\ \rm{nm}$ are consistent with previous reports \cite{Takahashi2012_P-Sr122, Kim2010}.
\begin{table}[b]
	\caption{\label{tab:sample properties}Properties of samples we measured. 
	$T_{\rm c}$ was defined by the criteria of $\rho_1=0.8\rho_n$.
	$\lambda(0)$ and $A$ were determined by fitting the data with $\lambda^{-2}(T)=\lambda^{-2}(0)[1-A(T/T_{\rm c})^{2}]$ upto $0.3T_{\rm c}$.
	The initial slope, ${\rm d}B_{\rm c2}/{\rm d}T|_{T_{\rm c}}$, was determined by $T_{\rm c}(B)$ obtained from $\rho_1(T,B)$.
	}
	\begin{tabular}{ccccc}\hline\hline
		\ \ batch\ \ &\ \ $T_{\rm c}$ (K)\ \ &\ \ $\lambda(0)$ (nm)\ \ &\ \ $A$\ \ &\ \ ${\rm d}B_{\rm c2}^{\parallel c}/{\rm d}T|_{T_{\rm c}}$ (T/K)\ \ \\\hline
		\#1&$14.5$&537&1.58&$-5.3\pm0.6$\\
		\#2&$14.6$&506&1.66&$-5.4\pm0.4$\\
		\#3&$14.5$&520&1.50&$-5.5\pm0.5$\\
		\#4&$14.6$&557&1.39&$-5.8\pm0.5$\\\hline\hline
	\end{tabular}
\end{table}
Deviations of $\lambda(0)$ mainly came from errors of the estimate of sample dimensions in the process to determine $\rho_{\rm dc}$ since we determined the magnitude of $Z_{\rm s}$ from $\rho_{\rm dc}$ directly.
Small variations of $T_{\rm c}$ within 1.5\% and good agreement in the superfluid-density fraction, $\lambda^2(0)/\lambda^2(T)$, shown in the inset of Fig. \ref{fig:lambdavsT} indicate that variations of physical properties due to the difference in composition were small.

\begin{figure}[h]
	\includegraphics[width=0.9\hsize]{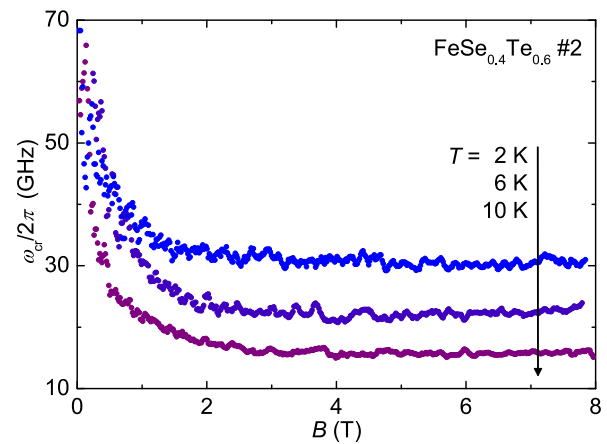}
	\caption{\label{fig:omega_crvsB}(Color online) The crossover frequency of FeSe$_{0.4}$Te$_{0.6}$ batch \#2 as a function of magnetic field measured at $T=2,6$, and 10 K.}
\end{figure}

Figure \ref{fig:omega_crvsB} shows that the crossover frequency, $\omega_{\rm{cr}}/2\pi$, decreased as $B$ and $T$ increased.
Such $B$- and $T$ dependence is consistent with the conventional understanding that increasing the driving force and thermal fluctuations weaken a pinning force, and similar behavior have been observed in other Fe-SCs \cite{Okada2012_Li111, Okada2013_P-Li111, Takahashi2012_P-Sr122}.
The observed value of $\omega_{\rm{cr}}(2\ {\rm K})/2\pi\gtrsim30\ \rm{GHz}$ is much larger than that of LaFeAsO$_{0.9}$F$_{0.1}$ ($\approx6\ \rm{GHz}$) \cite{Narduzzo2008} and of Li111 ($\approx3\ \rm{GHz}$) \cite{Okada2012_Li111}, suggesting that FeSe$_{0.4}$Te$_{0.6}$ has very strong pinning nature, which is quantitatively consistent with a large critical current density \cite{Si2013,Sun2013}.

\begin{figure}[b]
	\includegraphics[width=0.9\hsize]{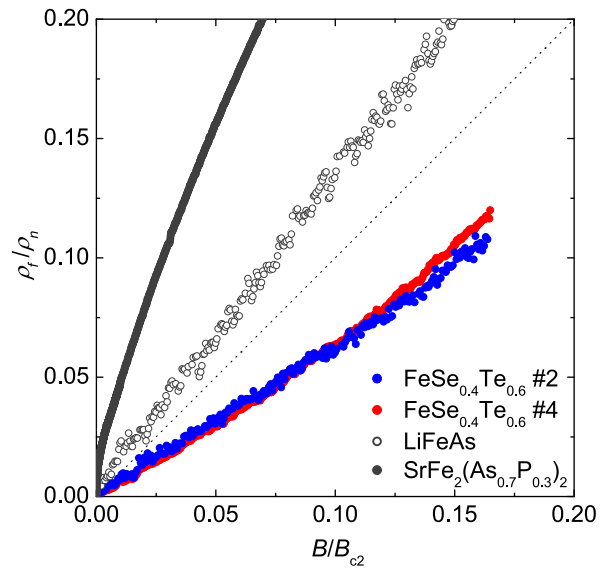}
	\caption{\label{fig:rho_fvsB}(Color online) The magnetic field dependence of the flux-flow resistivity of FeSe$_{0.4}$Te$_{0.6}$ batches \#2 (blue) and \#4 (red) measured at $T/T_{\rm c}\approx0.13$.
	For comparison, the same plots of Li111 (gray open, $T/T_{\rm c}\approx0.11$ \cite{Okada2012_Li111}), P-Sr122 (gray solid, $T/T_{\rm c}\approx0.08$ \cite{Takahashi2012_P-Sr122}), and B-S's prediction (dotted line) are also shown.}
\end{figure}

Figure \ref{fig:rho_fvsB} shows the $B$-dependence of the flux-flow resistivity measured at $T=2\ \rm{K}$.
The vertical axis is normalized by $\rho_{n}(T)$, and the horizontal axis is normalized by the upper critical field, $B_{\rm c2}(T)$.
The corresponding plots for fairly clean Fe-SCs are also shown for comparison.
Using the value of $B_{\rm c2}=48\ \rm{T}$ \cite{Khim2010}, the gradient of $\rho_f(B)$ found to be $\alpha^{\rm{FeSe_{0.4}Te_{0.6}}}\approx0.66$.
Here, $B_{\rm c2}$ value should be considered carefully because it relates to $\alpha$ directly.
In the B-S model \cite{Bardeen1965}, $B_{\rm c2}$ is defined by the critical field in the orbital limit where vortex cores occupy the entire sample, i.e., $B_{\rm c2}=B_{\rm c2}^{\rm orb.}$.
However, it is difficult to to determine $B_{\rm c2}^{\rm orb.}$ of Fe-SCs because of the multiple-band nature.
Moreover, several experiments under high magnetic fields \cite{Kida2009,Lei2010,Khim2010} reported that observed $B_{\rm c2}(T)$s of FeSe$_{1-x}$Te$_x$ system are strongly affected by the Pauli paramagnetic effect, i.e., $B_{\rm c2}<B_{\rm c2}^{\rm orb.}$.
This condition also makes it difficult to measure $B_{\rm c2}^{\rm orb.}$ directly.
To obtain $B_{\rm c2}^{\rm orb.}$ in FeSe$_{1-x}$Te$_x$ system, Khim {\it et al.} \cite{Khim2010} and Lei {\it et al.} \cite{Lei2010} fitted the data measured under high magnetic fields with a WHH formula including the Pauli-limiting effect, and reported $(B_{\rm c2}^{\rm orb.}(0),\ {\rm d}B_{\rm c2}^{\parallel c}/{\rm d}T|_{T_{\rm c}})=(56.5\ {\rm T},\ -5.6\ {\rm T/K})$ and $(57.9\ {\rm T},\ -5.8\ {\rm T/K})$, respectively.
These initial slopes, ${\rm d}B_{\rm c2}^{\parallel c}/{\rm d}T|_{T_{\rm c}}$, agree well with our data listed in Table \ref{tab:sample properties}.
Using these $B_{\rm c2}^{\rm orb.}$ values to normalize the horizontal axis of Fig. \ref{fig:rho_fvsB} yields $\alpha\approx0.78$, which are still smaller than unity.
Thus, we consider this small gradient to be an essential characteristic of FeSe$_{0.4}$Te$_{0.6}$.
$\alpha$ smaller than unity is considerably different from previously reported values for other Fe-SCs, i.e., $\alpha^{\rm{Co\mathchar`-Na111}}\approx1$, $\alpha^{\rm Li111}\approx1.4$, $\alpha^{\rm{P\mathchar`-Sr122}}\approx3.3$, and $\alpha^{\rm{P\mathchar`-Ba122}}\approx3.2$ \cite{Okada2012_Li111,Okada2013_P-Li111,Okada2013_Co-Na111,Takahashi2012_P-Sr122,Okada2014_P-Ba122}.
Previous flux-flow studies on cuprates, two-band systems, and Fe-SCs have shown that (i) the sign-change of the gap function is not essential for $\rho_f(B)$ \cite{Tsuchiya2001,Matsuda2002,Okada2012_Li111}, (i\hspace{-0.1em}i) the multiple-gap nature results in $\alpha>1$ \cite{Shibata2003,Akutagawa2008,Goryo2005,Takahashi2012_P-Sr122,Okada2014_P-Ba122}, and (i\hspace{-0.1em}i\hspace{-0.1em}i) the anisotropic gap function also results in $\alpha>1$ \cite{Kopnin1997,Kambe1999,Tsuchiya2001,Matsuda2002,Takaki2002,Okada2012_Li111,Okada2013_P-Li111,Okada2013_Co-Na111,Takahashi2012_P-Sr122,Okada2014_P-Ba122}.
Therefore, the observed small gradient, $\alpha^{\rm{FeSe_{0.4}Te_{0.6}}}<1$, is hard to be understood by these features.
\begin{figure*}[t]
	\includegraphics[width=0.9\hsize]{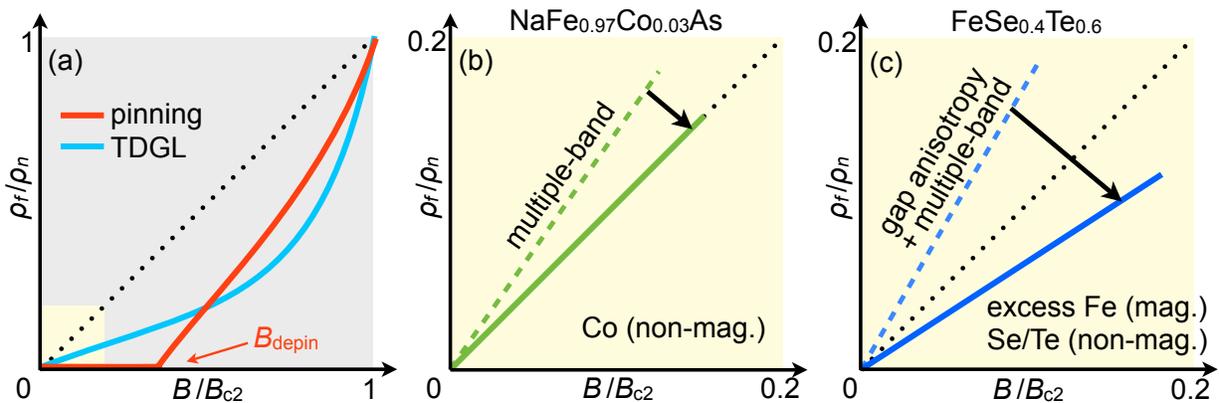}
	\caption{\label{fig:Schematics}(Color online) (a) A schematic of $\rho_f/\rho_n$ as a function of $B/B_{\rm c2}$ based on the pinning-induced back-flow model (red, \cite{Wang1992}) and on the TDGL theory (sky blue, \cite{Hu1973}).
	Panels (b) and (c) show $\rho_f(B\ll B_{\rm c2})$ of Co-Na111 and of FeSe$_{0.4}$Te$_{0.6}$, respectively.
	The dashed and solid lines are expected behaviors without the intrinsic back-flow current and experimentally-observed behaviors.
	The dotted lines in all panels are B-S's prediction.}
\end{figure*}

A possible explanation for the small $\alpha$ is effects of disorder.
The obtained results of (i) the large residual dc resistivity, (i\hspace{-0.1em}i) the $T^2$-dependence of the superfluid density, and (i\hspace{-0.1em}i\hspace{-0.1em}i) the large crossover frequency indicate that FeSe$_{0.4}$Te$_{0.6}$ contains a large amount of disorder, even in single crystals.
This characteristic is in sharp contrast to that of fairly clean Fe-SCs such as Li111, P-Sr122, and P-Ba122.
Thus, we consider that this highly-disordered nature of FeSe$_{0.4}$Te$_{0.6}$ induced the observed small $\alpha^{\rm{FeSe_{0.4}Te_{0.6}}}$.
Actually, similar small gradients (or corresponding steep enhancements just below $B_{\rm c2}$) of $\rho_{f}(B)$ have been observed experimentally in superconducting alloys with high concentration of disorder, such as Nb-Ta \cite{Kim1965,Pedersen1973}, Ti-V \cite{Kim1965}, Al-In \cite{Pedersen1973}, and Pb-In \cite{Fogel1973} systems.
A well-known role of disorder in SCs is to introduce pinning centers.
If one measures $\rho_f(B)$ of SCs with strong pinning by using a dc technique, $\rho_f(B)$ should be non-dissipative below the depinning field, $B_{\rm depin}=F_{\rm pin}/j$ ($F_{\rm pin}$: the pinning force) \cite{TinkhamTEXT, Wang1992}.
A schematic image of this behavior is shown in Fig. \ref{fig:Schematics}(a).
This is because that the vortex pinning disturbs a vortex motion and reproduces a back-flow current in the vicinity of the vortex, making the electric field induced inside the vortex core, ${\bm E}_{\rm core}$, suppressed \cite{TinkhamTEXT}.
However, the flux-flow resistivity we obtained does not suffer from the vortex pinning since we measured both of the reactive- and the resistive part of $Z_{\rm s}$ with a microwave frequency and derived $\rho_f(B)$ from those data.
In fact, our $\rho_f(B)$ data is clearly different from that affected by the back-flow current due to the vortex pinning.
Therefore, $\alpha^{\rm{FeSe_{0.4}Te_{0.6}}}<1$ should be caused by another effect of disorder.
Theoretically, $\rho_f(B)$ with small $\alpha$ was reproduced by the time-dependent Ginzburg-Landau (TDGL) equation for gapless SCs with pair-breaking due to magnetic impurities as shown in Fig. \ref{fig:Schematics}(a).
Here we describe a brief summary of theoretical studies related to the TDGL equation for gapless SCs below.
The first attempt to extend the GL theory to time-dependent situation and to describe energy dissipations in the mixed state by this scheme was conducted by Schmid \cite{Schmid1966}, and further extensions were achieved by some authors \cite{Caroli1967,Takayama1970}.
Complete sets of the TDGL equation for SCs with strong- and weak pair-breaking due to magnetic impurities were microscopically derived by Gor'kov-Eliashberg \cite{Gorkov1968} and by Eliashberg \cite{Eliashberg1969}, respectively.
Combining the complete sets of TDGL equation with the Maxwell equation leads to a differential equation for a gauge-invariant scalar- and vector potential, $\tilde{\varphi}\equiv\varphi+(\hbar/2e)\partial\chi/\partial t$ and $\tilde{\bm A}\equiv{\bm A}-(\hbar/2e)\nabla\chi$ ($\chi$ is the phase of the superconducting order parameter: $\Delta(\bm r)=\Delta_0f(\bm r){\rm e}^{{\rm i}\chi(\bm r)}$ where $\Delta_0$ is the gap size far away from the vortex core), as
\begin{equation}
	\left(\nabla^2+\frac{f^2}{\mu_0\sigma_nD\lambda^2}\right)\tilde{\varphi}=-\nabla\cdot\frac{\partial\tilde{\bm A}}{\partial t}+\frac{\partial\rho}{\partial t},
\end{equation}
where $\sigma_n=1/\rho_n$ is the normal-state conductivity and $D=v_{\rm F}^2\tau/3$ is a diffusion constant of electron.
Then, a screening length for $\tilde{\varphi}$ would be naturally introduced as $\zeta\equiv\lambda\sqrt{\mu_0\sigma_nD}$. 
Thompson and Hu \cite{Thompson1971,Hu1973} clarified that (i) the assumption in Refs. \cite{Bardeen1965,Schmid1966,Caroli1967} that uniform electric fields, ${\bm E}_{\rm core}={\bm B}\times{\bm v}_v$ (${\bm v}_v$: velocity of the vortex), are induced inside the vortex core holds only when $\zeta=\lambda$ and (i\hspace{-0.1em}i) non-uniform electric fields are induced when $\zeta\neq\lambda$ since local charges are different from those expected for the low-velocity Lorentz transformation of locations of vortices, ${\bm r}_i\rightarrow{\bm r}_i-{\bm v}_vt$.
According to their calculation, the total current is composed by the superfluid current constituting a vortex lattice ${\bm j}_s$, the transport current flowing through vortex cores uniformly,
\begin{equation}
	{\bm j}_t=\sigma_n\left(1+\frac{\xi^2}{2\zeta^2}\Braket{\left|\Delta\right|^2}\right)\Braket{\bm B}\times{\bm v}_v,
\end{equation}
where $\Braket{X}$ is the spatially averaged number of $X$, and the back-flow current distributing around each vortices with a dipole-like shape, ${\bm j}_b$.
We call ${\bm j}_b$ as the {\it intrinsic} back-flow current in this paper in order to distinguish it from the back-flow current caused by the vortex pinning mentioned previously.
The intrinsic back-flow current inside the vortex core is given by
\begin{equation}
	{\bm j}_b^{\rm in}=\sigma_n\left(1-\frac{\lambda^2}{\zeta^2}\right)[{\bm B}-\Braket{\bm B}]\times{\bm v}_v.
\end{equation}
${\bm j}_b^{\rm in}$ flows counter to ${\bm j}_t$ if $\zeta$ is smaller than $\lambda$ and becomes remarkable when the scattering time $\tau$ is small since $\lambda^2/\zeta^2=3m^*/\mu_0ne^2v_{\rm F}^2\tau^2$.
Simultaneously, the second term of ${\bm j}_t$ relating to a relaxation of the order parameter \cite{Tinkham1964} should be enhanced in order to meet the equation of continuity $\nabla\cdot{\bm j}+\partial\rho/\partial t=0$, and energy dissipations in the vortex core, ${\bm j}_t\cdot\Braket{\bm E}=\eta{\bm v}_v^2$ ($\eta$: the viscous-drag coefficient), should increase.
This indicates that the flux-flow resistivity, $\rho_f=\Phi_0B/\eta$, in highly-disordered system, where the intrinsic back-flow phenomenon is significant, becomes smaller than that predicted in the B-S model.
By using the microscopically-expected number of $\zeta=\xi/\sqrt{12}$, numerical calculations of the TDGL equations for gapless SCs with high concentration of magnetic impurities reported $\alpha$ to be 0.38 \cite{Kupriyanov1972} and 0.33 \cite{Hu1972}.
Therefore, $\alpha<1$ is a manifestation of the intrinsic back-flow phenomenon remarkable in highly-disordered SCs.
Returning to the case of FeSe$_{1-x}$Te$_x$, excess-Fe atoms are well-known to act as magnetic impurities \cite{Komiya2013, Zhang2009}.
Thus, it is expected that FeSe$_{1-x}$Te$_{x}$ with excess-Fe atoms behaves similarly to conventional SCs with paramagnetic impurities, and we consider that the observed small $\alpha$ of FeSe$_{0.4}$Te$_{0.6}$ also originates from the intrinsic back-flow phenomenon.
The magnetic vortex in multiple-band SCs is not understood even theoretically because of the complexity of the system.
Therefore, this first experimental observation of the intrinsic back-flow phenomenon in these SCs is highly significant.

Finally, we consider the difference between Co-Na111 and FeSe$_{0.4}$Te$_{0.6}$.
If the intrinsic back-flow effect is negligibly small, the gradient $\alpha$ of these materials should be larger than unity because Co-Na111 has multiple bands with almost isotropic electronic states \cite{Liu2011} and FeSe$_{0.4}$Te$_{0.6}$ has multiple bands with anisotropic nodeless gaps \cite{Zeng2010,Okazaki2012}.
Practically, the intrinsic back-flow current of these materials is not negligible, and we observed that the $\alpha$ values of these materials were suppressed.
These behaviors are shown in Figs. \ref{fig:Schematics}(b) and (c) as dashed lines (which correspond to the predicted behavior in the clean limit; without the intrinsic back-flow current) and solid lines (which correspond to behavior we measured; with the intrinsic back-flow current).
Although both Co-Na111 and FeSe$_{0.4}$Te$_{0.6}$ exhibited gapless superconductivity, different $\alpha$ values were observed for the two materials: $\alpha^{\rm{Co\mathchar`-Na111}}\approx1$ and $\alpha^{\rm FeSe_{0.4}Te_{0.4}}\approx0.66$.
This difference could be attributed to the differences in the type and amount of impurities.
In Ref. \cite{Hu1973}, the $\alpha$ value of conventional SCs was calculated as a function of the spin-flip scattering rate $\tau_{\rm s}^{-1}$ and the total scattering rate $\tau_1^{-1}$: if the pair-breaking by spin-flip scattering is not too strong, $\alpha$ could be larger than unity when the total scattering rate is similar to that resulting from magnetic impurities ($\tau_1^{-1}\approx\tau_{\rm s}^{-1}$) and becomes less than unity as the scattering rate by non-magnetic impurities becomes large ($\tau_1^{-1}\gg\tau_{\rm s}^{-1}$).
Although it is not clear whether these predictions are quantitatively valid at present, a similar trend is expected for multiple-band SCs.
Recent scanning tunneling microscopy/spectroscopy studies on NaFe$_{0.97-y}$Co$_{0.03}${\it T}$_{y}$As ({\it T}=Cu, Mn) showed that Co atoms are non-magnetic or weak-magnetic impurities \cite{Yang2013}, suggesting that the condition $\tau_1^{-1}\approx\tau_{\rm s}^{-1}$ is satisfied for Co-Na111.
In contrast, excess-Fe atoms (i.e., corresponding to atomic concentrations below 4\%) and doped Se/Te atoms (Se 37\%, Te 63\%) in FeSe$_{0.4}$Te$_{0.6}$ behaved as magnetic impurities and non-magnetic impurities, respectively.
This finding most likely corresponds to the condition $\tau_1^{-1}\gg\tau_{\rm s}^{-1}$.
Therefore, the strongly suppressed $\alpha^{\rm FeSe_{0.4}Te_{0.6}}$ may be attributed to the combination of a small amount of magnetic impurities and a large amount of non-magnetic impurities in contrast to the weak suppression of $\alpha^{\rm{Co\mathchar`-Na111}}$ by a small amount of non-magnetic impurities (Co 3\%).
Although we do not as yet understand the explicit relationship between the amount of disorder of a sample and its $\alpha$ value, this relationship could be clarified by performing more systematic studies of $\rho_f(B)$ for FeSe$_{1-x}$Te$_{x}$ with different amounts of excess-Fe atoms and/or that of Co-Na111 containing magnetic impurities, such as Mn.
\section{I\hspace{-0.1em}V. Conclusions}
We measured the microwave surface impedance of FeSe$_{0.4}$Te$_{0.6}$ single crystals both in the zero-field limit and under finite magnetic fields.
The superfluid density measured under the zero-external field behaved as $\lambda^{-2}(T)-\lambda^{-2}(0)\propto (T/T_{\rm c})^2$, indicating a strong pair-breaking effect in this material.
The data obtained under finite magnetic fields showed that $\omega_{\rm cr}/2\pi$ for FeSe$_{0.4}$Te$_{0.6}$ was much larger than that of LiFeAs and of LaFeAsO$_{0.9}$F$_{0.1}$, suggesting a strong pinning.
The gradient of $\rho_f(B\ll B_{\rm c2})$ was $\alpha^{\rm FeSe_{0.4}Te_{0.6}}\approx0.66$ with $B_{\rm c2}(0)=48\ {\rm T}$, which is considerably smaller than that of other Fe-SCs ($\alpha\geq1$).
We attributed this small $\alpha$ to the intrinsic back-flow current remarkable in highly-disordered materials, which should provide valuable information on the understanding of vortices in multiple-band SCs.

\begin{acknowledgments}
We thank Yusuke Masaki, Noriyuki Kurosawa, Yoichi Higashi, and Dr. Yuki Nagai for the fruitful discussions on theoretical aspects.
We also thank Dr. Seiki Komiya and Dr. Ichiro Tsukada for showing us their unpublished data and the valuable comments.
This work was partially supported by Strategic International Collaborative Research Program (SICORP), Japan Science and Technology Agency.
\end{acknowledgments}

\bibliography{141026Bibliography.bib}
\end{document}